\documentclass[conference]{IEEEtran}
\IEEEoverridecommandlockouts
\usepackage{cite}
\usepackage{amsmath,amssymb,amsfonts,orcidlink}
\usepackage{algorithmic}
\usepackage{graphicx}
\usepackage{textcomp}
\usepackage{xcolor}
\usepackage{booktabs}
\usepackage{array}
\usepackage{hyperref}

\makeatletter \newcommand{\linebreakand}{ \end{@IEEEauthorhalign} \hfill\mbox{}\par \mbox{}\hfill\begin{@IEEEauthorhalign} } \makeatother

\begin{document}

\title{Managing Iterative Hybrid Quantum--Classical Optimization as a First-Class Scientific Workflow}

\author{\IEEEauthorblockN{Giuliana Siddi Moreau \orcidlink{0000-0002-0945-6915}}
\IEEEauthorblockA{\textit{CRS4 - Center for  } \\
\textit{Advanced Studies, Research } \\
\textit{and Development in Sardinia}\\
Pula, Italy \\
julie@crs4.it}
\and
\IEEEauthorblockN{Maria Laura Clemente}
\IEEEauthorblockA{\textit{CRS4 - Center for  } \\
\textit{Advanced Studies, Research } \\
\textit{and Development in Sardinia}\\
Pula, Italy \\
clem@crs4.it}
\and
\IEEEauthorblockN{Lorenzo Pisani}
\IEEEauthorblockA{\textit{CRS4 - Center for  } \\
\textit{Advanced Studies, Research } \\
\textit{and Development in Sardinia}\\
Pula, Italy \\
pisani@crs4.it}
\linebreakand
\IEEEauthorblockN{Manuela Profir}
\IEEEauthorblockA{\textit{CRS4 - Center for  } \\
\textit{Advanced Studies, Research } \\
\textit{and Development in Sardinia}\\
Pula, Italy \\
manuela@crs4.it}
\and
\IEEEauthorblockN{Marco Pinna}
\IEEEauthorblockA{\textit{CRS4 - Center for  } \\
\textit{Advanced Studies, Research } \\
\textit{and Development in Sardinia}\\
Pula, Italy \\
marco.pinna@crs4.it}
\and
\IEEEauthorblockN{Marco Moro}
\IEEEauthorblockA{\textit{CRS4 - Center for  } \\
\textit{Advanced Studies, Research } \\
\textit{and Development in Sardinia}\\
Pula, Italy \\
mmoro@crs4.it}
\and
\IEEEauthorblockN{Lidia Leoni}
\IEEEauthorblockA{\textit{CRS4 - Center for  } \\
\textit{Advanced Studies, Research } \\
\textit{and Development in Sardinia}\\
Pula, Italy \\
lidia.leoni@crs4.it}
}

\maketitle

\begin{abstract}
Today's Quantum Processing Units (QPUs) are too small and too noisy to solve large combinatorial optimization problems directly, so practical hybrid solvers split a problem into pieces and iterate a decompose--solve--aggregate loop over whatever backends are available: classical heuristics, simulators, emulators, or a QPU. In practice, the loop is a driver script. It sits on top of the quantum-HPC middleware, handling task generation, provenance, recovery, and portability. Instead, we treat the loop as a scientific workflow and ask what a workflow layer adds to a generic workflow management system and QPU-sharing middleware. Two decomposition patterns from real applications, iterative consensus (ADMM) and hierarchical partitioning, turn out to stress the orchestration layer very differently: over 120 managed runs, orchestration took 78.4\% of end-to-end time for the iterative pattern, almost all of it in a per-round barrier, but only 6.2\% for the hierarchical one. Our workflow model adds four things a generic engine does not have: a termination predicate residing in the task graph that reads the previous round's residuals, subproblem-level recovery with warm-start and quorum-deferred aggregation, failover from a QPU to a classical replica within a round, and a provenance schema that describes CPUs, simulators and QPUs with the same fields, including the shot budget included. We report the cost of each layer on our engine, show that speculative re-execution enable runs to complete under injected failures that stall an unmanaged driver. We also use the same provenance to give per-device latency tails across simulators, emulators and IQM QPUs.
\end{abstract}


\section{Introduction}
Quantum processing units (QPUs) are starting to show up in high-performance computing (HPC) centres alongside to CPUs and GPUs \cite{ornl,riken,cesga}, and there is now a fair amount of middleware to schedule them: Slurm extensions that expose a QPU as a resource, runtimes that dispatch circuits across devices, and sharing strategies that decide how several hybrid jobs divide one scarce device \cite{sitdikov2025qrm,esposito2025slurmhet,cipollini2026share,shehata2025bridging}. 
This paper is about what sits on top of all that. An application-level loop must decide what to submit when a round is finished, what to do when a submission fails, and what information to record so that the run can be explained later.

Our applications are combinatorial optimization problems written as binary quadratic models (BQMs), which are consumed by annealers and gate-model heuristics. Realistic instances are too large for one device, so hybrid solvers divide the model into subproblems, solve each on a backend, and combine the results over several rounds.  QBSolv and D-Wave Hybrid work this way \cite{qbsolv,dwavehybrid}, and so do operator-splitting methods built on ADMM \cite{SiddiMoreauQCE2026}. We are not interested here in any particular decomposition algorithm, only in the machinery that runs the loop.

That machinery is usually a driver script. It submits subproblems to a batch system or a cloud queue, waits, and stitches the answers together. There is no explicit task graph, little provenance, retries are done by hand if at all, and the script is welded to one site's scheduler. None of this matters much while every backend is a fast local CPU. It starts to matter when one backend is a QPU, because a QPU is scarce, remotely queued, drifts with calibration, and fails now and then, so its use has to be planned, recorded, retried and reproduced.

The obvious fix is to write the loop in a generic workflow management system (WMS). Parsl, Dask and PyCOMPSs all support task graphs that grow at runtime, and Pegasus and Covalent give provenance and retries out of the box \cite{parsl,dask2015,pycompss2017,pegasus,covalent}.The loop can certainly be written in any of those WMS, and our execution layer is built to run on them. Our point is a narrower one. Several things the loop needs are not properties of a task graph, so no task-graph engine provides them. 
The termination test has to read a residual that the previous aggregation computed. A failed block should be restarted from its value in the previous round, not from zero. Aggregation should progress once enough blocks are back, so that one stuck QPU submission does not hold the whole round. A single block should be able to move from a QPU to a classical quantum simulator without leaving a hole in the round's provenance. And the provenance record should describe a CPU, a simulator and a QPU with the same fields, shot budget included. These are properties of the decompose--solve--aggregate loop itself. Providing them once, for every decomposition pattern, is what this paper contributes.

Concretely, the paper features four contributions; Table~\ref{tab:contribmap} says where each one lives.
\begin{enumerate}
\item A workload characterization of two decomposition patterns from real applications, iterative consensus and hierarchical partitioning, as runtime-expanded task graphs with different barrier structure, together with measurements showing that their orchestration demands differ by an order of magnitude (Sec.~\ref{sec:workload}, Sec.~\ref{sec:rq1}).
\item A workflow model that adds to a generic engine a feedback-driven termination node, subproblem-level recovery with warm-start and quorum-deferred aggregation, within-round backend failover, and run-scope reschedule from checkpoint (Sec.~\ref{sec:model}).
\item A backend-agnostic provenance schema for hybrid runs (Table~\ref{tab:schema}) that records device and shot budget per subproblem; every number in this paper comes out of it (Sec.~\ref{sec:model}, Sec.~\ref{sec:rq3}).
\item An evaluation of what the layer costs, how it behaves under injected failures, and what latency tails each device shows, across simulators, emulators and IQM QPUs (Sec.~\ref{sec:results}).
\end{enumerate}

The model can also express nested patterns (partition first, then iterate inside each branch). Sec.~\ref{sec:patterns} explains how; we have not evaluated them yet.

\begin{table}[t]
\caption{Contributions mapped to sections, requirements and research questions.}
\label{tab:contribmap}
\centering
\small
\begin{tabular}{@{}p{2.7cm}p{1.7cm}p{1.0cm}p{1.6cm}@{}}
\toprule
\textbf{Contribution} & \textbf{Section} & \textbf{Req.} & \textbf{Evidence} \\
\midrule
Workload characterization & III, VII-A & R-A & RQ1, Table~\ref{tab:overheadpat} \\
Workflow model & IV & R-A, R-C & RQ2, Fig.~\ref{fig:resilience} \\
Provenance schema & IV-B, Table~\ref{tab:schema} & R-B & RQ3, Table~\ref{tab:backend} \\
Cost and resilience evaluation & VI, VII & R-A--R-D & RQ1--RQ3 \\
\bottomrule
\end{tabular}
\end{table}

\section{Background and Related Work}
\label{sec:related}

We follow the terminology proposed by the workflows community \cite{wfterminology2025}: a \emph{workflow} is a set of tasks with dependencies, a \emph{workflow management system} enacts it on resources, and a workflow is \emph{dynamic} when its task set is only known during enactment. Ours is dynamic twice over. How many rounds there are depends on a convergence test, and for adaptive decompositions the number and membership of subproblems in a round depend on what the previous round produced.

\paragraph{Generic workflow systems} Table~\ref{tab:wms} checks the loop's four requirements (Sec.~\ref{sec:workload}) against the systems one would normally reach for. Parsl, Dask and PyCOMPSs build the task graph at runtime from a Python program, so the dynamic fan-out of R-A is easy to express in all three; PyCOMPSs also targets HPC schedulers directly and has been used for hybrid quantum--classical pipelines \cite{parsl,dask2015,pycompss2017}. Pegasus and Covalent give task-level provenance and retries \cite{pegasus,covalent}. What none of them has is anything specific to this loop: a retry that restarts from the previous iterate instead of re-running the task, an aggregation that proceeds on a quorum, a failover that moves one task from a QPU to a classical replica without breaking the round's provenance, or a provenance record that knows what a shot budget is. Those are the rows where Table~\ref{tab:wms} shows a gap, and they are what Sec.~\ref{sec:model} adds. We are not competing with these engines for task scheduling; the executor interface in Sec.~\ref{sec:model} is meant to let any of them do the enactment.

\begin{table*}[t]
\caption{Requirements of the decompose--solve--aggregate loop against generic WMSs and quantum--HPC middleware. \checkmark: provided; $\circ$: expressible by the user but not provided; --: not addressed. The last column is the layer proposed here, which targets the generic WMSs as executors.}
\label{tab:wms}
\centering
\small
\begin{tabular}{@{}p{6.0cm}>{\centering\arraybackslash}p{2.2cm}>{\centering\arraybackslash}p{2.0cm}>{\centering\arraybackslash}p{2.0cm}>{\centering\arraybackslash}p{1.8cm}@{}}
\toprule
\textbf{Requirement} & \textbf{Parsl, Dask, PyCOMPSs} & \textbf{Pegasus, Covalent} & \textbf{Q--HPC middleware} & \textbf{This layer} \\
\midrule
Runtime-expanded fan-out (R-A) & \checkmark & $\circ$ & -- & \checkmark \\
Feedback-driven termination node (R-A) & $\circ$ & $\circ$ & -- & \checkmark \\
Per-task provenance (R-B) & $\circ$ & \checkmark & -- & \checkmark \\
Device- and shot-stamped provenance (R-B) & -- & -- & $\circ$ & \checkmark \\
Task-level retry (R-C) & \checkmark & \checkmark & -- & \checkmark \\
Warm-start retry, quorum aggregation (R-C) & -- & -- & -- & \checkmark \\
QPU$\to$classical failover within a round (R-C) & -- & -- & $\circ$ & \checkmark \\
Multi-scheduler enactment (R-D) & \checkmark & \checkmark & \checkmark & via executors \\
QPU placement and sharing & -- & -- & \checkmark & delegated \\
\bottomrule
\end{tabular}
\end{table*}

\paragraph{Quantum--HPC middleware} HPC facilities expose QPUs through integrated testbeds, supercomputer co-execution and emulated infrastructures \cite{ornl,riken,cesga}. Resource-manager extensions make QPUs first-class schedulable resources with co-scheduling, heterogeneous launches and shared-device allocation \cite{sitdikov2025qrm,esposito2023hybrid,esposito2025slurmhet}. Cipollini et al.\ compare three strategies for sharing one QPU among hybrid jobs and quantify the trade-off between exclusive access, time-slicing and a second-level scheduler \cite{cipollini2026share}. Middleware architectures and portable runtimes provide late binding, vendor-neutral execution, observability and concurrent dispatch across CPUs, GPUs, simulators and QPUs \cite{saurabh2023arch,mantha2026pilotq,wennersteen2025usercentric,miniskar2026qiree,shehata2025bridging,seelam2026qcsc,faro2023middleware}. Cloud-native systems use Kubernetes or provider-level orchestration to manage quantum kernels and trade off queue time, availability and fidelity \cite{tejedor2026k8s,mahesh2025conqure,giortamis2024qonductor,pehlivanoglu2026qurator}. All of these decide where and when a circuit runs. Our layer sits above them and decides what to submit, whether a round is done, and what to do when a submission fails. The per-device latency tails we record (Sec.~\ref{sec:rq3}) are exactly the input a sharing policy like \cite{cipollini2026share} would want. Qurator is the closest of these systems, since it also models hybrid workloads as dynamic DAGs \cite{pehlivanoglu2026qurator}, but its graph changes because of circuit transformations and provider choices; ours changes because of repeated decomposition, aggregation and feedback.

\paragraph{The application class} qbsolv, dwave-hybrid and operator-splitting methods all implement decompose--solve--aggregate for annealers and gate-model devices \cite{qbsolv,dwavehybrid,SiddiMoreauQCE2026}. They are about decomposition algorithms and solution quality; for us they are the workload. The pattern is not limited to optimization, either. Fragment-based quantum chemistry, distributed circuit cutting and knitting \cite{tejedor2025cutting}, and domain-decomposed simulation all fan subproblems out to mixed backends and reconcile them, and they run into the same barrier costs, straggler sensitivity and cross-backend provenance questions we measure here. We evaluate on optimization simply because that is where we have production workloads.

\section{Hybrid Optimization as a Workflow}
\label{sec:workload}
We start with the workload itself, described without reference to any particular decomposition algorithm or solver, and then work out what it asks of a WMS.

\subsection{The Decompose--Solve--Aggregate Loop}
An outer loop maintains a global candidate assignment. Each iteration (i) decomposes the residual problem into a set of subproblems whose number and membership depend on the current state; (ii) dispatches the subproblems, which are mutually independent within the iteration, to a pool of interchangeable backends; and (iii) aggregates the returned samples into an updated global assignment and tests for convergence. The loop ends when no further improvement is found or a budget is exhausted.

\subsection{Decomposition Patterns and Fan-out}
\label{sec:patterns}
The skeleton above hides some real variety. Different applications decompose in structurally different ways, and a workflow layer should support those patterns under one engine rather than hard-code one of them. The two applications we orchestrate (Sec.~\ref{sec:eval}) use one pattern each, and, as Sec.~\ref{sec:rq1} shows, the two cost very different amounts to orchestrate.

\begin{figure*}[t]
\centering
\includegraphics[width=0.85\textwidth]{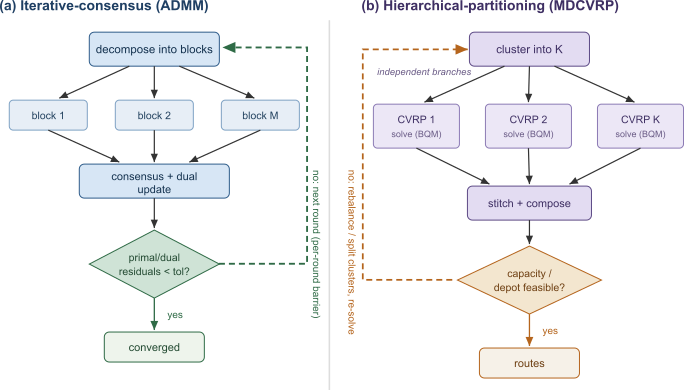}
\caption{Two decomposition/fan-out patterns expressed by the same workflow engine. (a) Iterative consensus (ADMM): the block fan-out recurs every round, closed by a primal/dual residual test and a per-round barrier. (b) Hierarchical partitioning (MDCVRP): a one-pass cluster-and-solve tree closed by a feasibility/stitching check whose failure rebalances or splits clusters and re-solves the affected branches.}
\label{fig:patterns}
\end{figure*}

Fig.~\ref{fig:patterns} shows both, expressed by the same engine through configurable decomposition and termination operators. In iterative consensus, based on the Alternating Direction Method of Multipliers (ADMM) and used by our renewable-energy-community application, each round distributes $M$ coupled block BQMs and performs consensus and dual updates before testing the primal and dual residuals. This requires repeated fan-out, a per-round barrier, and provenance of the residual trajectory. In hierarchical partitioning, used by our fleet-routing application for the Multi-Depot Capacitated Vehicle Routing Problem (MDCVRP), customers are clustered into independent single-depot CVRP subproblems whose solutions are stitched and checked for feasibility; failed clusters are rebalanced, split and re-solved, producing a bounded stitch-and-refine loop rather than a repeated global barrier. The two can also be nested, for instance by solving each hierarchical branch with ADMM, which gives two levels of fan-out sharing one set of provenance and fault-tolerance mechanisms. The engine can express this; we have not evaluated it in this paper. Table~\ref{tab:patterns} summarizes the orchestration demand of each.

\begin{table}[t]
\caption{Decomposition patterns supported by the same workflow engine, and the orchestration demands each places.}
\label{tab:patterns}
\centering
\small
\begin{tabular}{@{}p{1.7cm}p{2.3cm}p{3.2cm}@{}}
\toprule
\textbf{Pattern} & \textbf{Fan-out structure} & \textbf{Orchestration demand} \\
\midrule
Iterative consensus (ADMM) & repeated $M$-block fan-out + consensus/dual update & feedback-driven graph; residual provenance; per-round barrier \\
Hierarchical partitioning & tree of $K$ branches + stitch-and-refine feedback & partition/compose tasks; feasibility check; rebalance/split re-solve \\
Nested (partition $\to$ per-branch iterative) & two-level fan-out & hierarchical scheduling and provenance (not evaluated here) \\
\bottomrule
\end{tabular}
\end{table}

\subsection{Why a Static Task Graph Does Not Fit}
The fan-out of each iteration depends on the data, so the task graph cannot be drawn in advance. Tasks are interchangeable across very different hardware: the same subproblem may run on a CPU, a simulator or a QPU, and each has its own latency, cost and failure rate. Backends are stochastic samplers, so results come back partial and noisy, and the pipeline has to cope with late or failed samples without stalling the round or biasing the aggregate. And if a run is ever to be replayed, every intermediate state has to be recorded well enough to do so.

\subsection{Requirements}
Four requirements follow. \textbf{R-A}: dynamic task generation with feedback-aware termination. \textbf{R-B}: per-subproblem provenance across heterogeneous backends. \textbf{R-C}: idempotent, retryable subproblems with backend-level failover. \textbf{R-D}: portability across WMSs and schedulers without rewriting the loop. Table~\ref{tab:wms} shows which of these a generic WMS already meets.

\section{Workflow Model and Orchestrator}
\label{sec:model}
\begin{figure*}[t]
\centering
\includegraphics[width=0.80\linewidth]{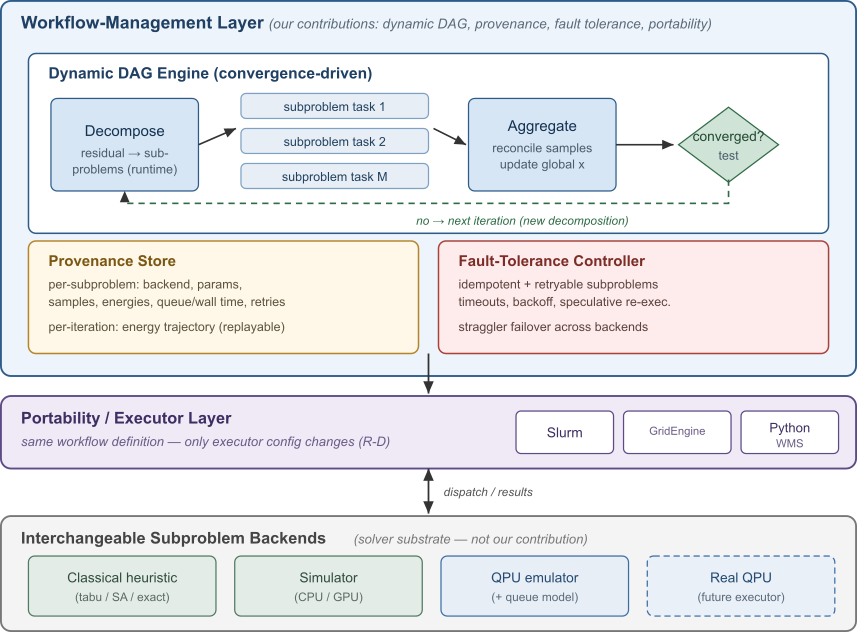}
\caption{The orchestrator as a workflow-management layer above interchangeable execution backends and enactment engines. The contributions are the dynamic-graph semantics, provenance, fault tolerance and portability, not the underlying solver or task scheduler.}
\label{fig:arch}
\end{figure*}

Fig.~\ref{fig:arch} shows the overall shape. The orchestrator sits between the decomposition loop and a pool of interchangeable backends. It owns the dynamic-graph semantics, the provenance, the fault tolerance and the portability; both the solver and the engine that actually enacts tasks are replaceable parts.

\subsection{Dynamic Graph with Feedback-Driven Termination}
A run is a task graph that grows one iteration at a time. An iteration node calls the decomposition operator, which emits however many subproblem task nodes the current state calls for. An aggregation node collects their results, updates the global state, and emits either the next iteration node or a terminal node. The convergence test is itself a node: it reads the residuals the aggregation node wrote and decides whether the graph grows. Because the decomposition operator and the aggregation/termination predicate are parameters, iterative consensus, hierarchical partitioning and nested patterns are three configurations of one model, not three workflows (R-A). A generic WMS can host this graph without difficulty. What it does not provide is the termination node, or the recovery behaviour described below.

\subsection{Provenance Schema for Hybrid Runs}
\label{sec:schema}
Table~\ref{tab:schema} lists the schema. Each subproblem task writes a \texttt{BlockRecord}, each iteration an \texttt{IterationRecord}, each run a \texttt{RunRecord}. The fields are the same whether the backend was a CPU, a simulator or a QPU; the executor fills in the device descriptor and shot budget at solve time. This is what lets a hybrid run be reproduced and debugged (R-B), and it is the only source of the numbers in Sec.~\ref{sec:results}. The records can be queried in place or exported for artifact deposit.

\begin{table}[t]
\caption{Provenance schema. Every measurement in Sec.~\ref{sec:results} is computed from these fields.}
\label{tab:schema}
\centering
\small
\begin{tabular}{@{}p{2.6cm}p{2.6cm}p{2.0cm}@{}}
\toprule
\textbf{Field} & \textbf{Purpose} & \textbf{Consumer} \\
\midrule
\multicolumn{3}{@{}l}{\texttt{BlockRecord} (one per subproblem attempt)} \\
input fingerprint & identify the sub-BQM solved & replay, RQ1 \\
pattern facet & which decomposition pattern emitted it & Table~\ref{tab:overheadpat} \\
backend descriptor: provider, type, device, shots & attribute latency and cost to a device & Table~\ref{tab:backend}, sharing policy \\
solver parameters & reproduce the solve & replay \\
attempt number, retry reason & separate first-attempt from recovery & Fig.~\ref{fig:resilience}b \\
queue, execute, transfer times & decompose latency & RQ1, RQ3 \\
returned samples, energies & aggregation input & solver, RQ2 quorum \\
\multicolumn{3}{@{}l}{\texttt{IterationRecord} (one per round)} \\
primal/dual residuals & termination predicate & Fig.~\ref{fig:admm} \\
decomposition decisions & which blocks, which sizes & replay \\
barrier / stitch wait & orchestration cost per round & Table~\ref{tab:overheadpat} \\
quorum reached, blocks warm-started & recovery actions taken & RQ2 \\
\multicolumn{3}{@{}l}{\texttt{RunRecord} (one per run)} \\
termination class & converged / degraded / invalid & RQ2 \\
reschedule count, checkpoint iteration & run-scope recovery & RQ2 \\
executor target & scheduler / WMS used & R-D \\
\bottomrule
\end{tabular}
\end{table}

\subsection{Fault Tolerance and Straggler Mitigation}
Every subproblem is idempotent. It can be resubmitted freely, because aggregation only ever consumes validated, completed samples. The fault-tolerance controller applies per-backend timeouts, retries with backoff, and speculative re-execution of stragglers on another backend, so a slow or rejected QPU job is quietly covered by a classical replica and the iteration carries on (R-C). Two of its policies are specific to this loop, and they are the ones a generic task-level retry does not give you. First, aggregation waits for a \emph{quorum} of the iteration's subproblems rather than for all of them, so one stuck block cannot stall the round. Second, an invalid block is \emph{warm-started} from its value in the previous iteration instead of being zeroed or recomputed from scratch, which keeps the global iterate well-conditioned.

There is also recovery at the level of a whole run. A termination classifier labels every completed run as converged, degraded-but-feasible, or invalid (empty, non-finite energy, infeasible, or every block failed). If a run ends invalid, a reschedule controller runs it again, with bounded retries and exponential backoff, resuming from a per-iteration checkpoint of the solver state and optionally switching to a more reliable backend. How many reschedules it took to reach a valid termination goes into the \texttt{RunRecord}.

\subsection{Portability Layer}
The loop is written once, against an executor interface. Concrete executors map subproblem tasks onto a Slurm allocation, a Grid Engine job, or a Python task engine running inside a single allocation, and switching between them touches none of the decomposition, aggregation, provenance or fault-tolerance code (R-D). The same interface is where a generic WMS would plug in as the enactment engine, and where a remote-QPU executor adds embedding, queue handling and shot budgeting.

\section{HPC as a Quantum-Preparation Testbed}
\label{sec:testbed}
Fault-tolerant QPUs are not widely available, so we built and validated the workflow on HPC. We would argue this is the right place to do it, not a stopgap. Only one operation in the loop, solving a subproblem, is tied to a backend at all; decomposition, aggregation, provenance capture, fault handling and scheduling are not. To the loop a QPU is just a sampler that returns low-energy candidates, and a classical sampler can stand in for it during development. HPC parallelism rehearses the same fan-out, dependencies and aggregation contention at sizes no QPU can reach yet, and lets us inject the failures and latencies a QPU queue will produce.

We group the capabilities a workflow layer needs into four \emph{readiness levels}, none of which depends on how mature the hardware is:
\begin{itemize}
\item \emph{L1 Structural}: decomposition, fan-out and aggregation are expressed as a dynamic, feedback-driven graph.
\item \emph{L2 Recorded}: every subproblem and iteration is captured in backend-agnostic provenance sufficient to replay the run.
\item \emph{L3 Resilient}: subproblems are idempotent and retryable, with failover and straggler mitigation across heterogeneous backends.
\item \emph{L4 Portable and quantum-aware}: the same workflow runs across schedulers, and a QPU executor adds embedding, queue handling and shot budgeting without changing the loop.
\end{itemize}
L1 to L3 can be reached and measured on HPC today. L4 is exercised here partly on real IQM QPUs and otherwise against simulators, emulators and a queue model. The levels organize the evaluation; they are not a validated metric.

\section{Evaluation Methodology}
\label{sec:eval}
The evaluation is about orchestration, not solution quality. How good the decomposition solvers are is a separate question, addressed elsewhere.

As far as the platform is concerned, the experiments run on an HPC cluster of 124 nodes (dual Intel Xeon Gold 6226R, 2.90 GHz, 16 cores per socket; 3,968 cores), 384 GB RAM per node, 100 Gb/s InfiniBand EDR, scheduled with Grid Engine, together with a Slurm-managed virtual-machine cluster. Subproblem samplers include VQE, QAOA and digitized counterdiabatic quantum optimization (DCQO) implementations; backends include state-vector and matrix-product-state simulators, noise-aware emulators, and gate-model IQM QPUs accessed over their cloud queue.

 Regarding workloads, we feed the workflow BQM instances from two application domains. Each has its own front-end; everything downstream is shared.
\begin{itemize}
\item \emph{Fleet / vehicle routing} (CVRP and TSP variants), formulated as in \cite{vrpqubo}: binary variables $x_{v,i,t}$ encode that vehicle $v$ visits node $i$ at step $t$; one-hot and capacity constraints enter as quadratic penalties, giving dense, tightly constrained QUBOs. The multi-depot variant uses hierarchical partitioning: customers are clustered into single-depot CVRP subproblems, solved independently and stitched, with a feasibility check that rebalances or splits infeasible clusters and re-solves the affected branches.
\item \emph{Renewable energy communities} (member-to-community allocation and energy-sharing configuration) \cite{SiddiMoreauQCE2026}: binary variables assign prosumers to communities and toggle shared-asset decisions; penalty terms enforce per-interval energy balance and capacity, giving constraint-rich, strongly coupled BQMs. It uses iterative consensus (Alternating Direction Method of Multipliers - ADMM).
\end{itemize}

The research questions underlying the study can be summarized as follows: \\
\textbf{RQ1 (cost)}: what does the management layer cost as a fraction of end-to-end time, and how does that cost differ between decomposition patterns?\\
\textbf{RQ2 (resilience)}: under injected backend failures and stragglers, does the workflow complete, and at what overhead, with and without speculative re-execution?\\
\textbf{RQ3 (portability view)}: which sampler actually ran each subproblem, at what latency tail and shot cost, across simulators, emulators and QPUs?

\paragraph{Protocol} Solver and instances are held fixed, so any difference is due to the workflow layer. We define orchestration time as client-side end-to-end time minus the backend solve times recorded in the \texttt{BlockRecord}s. It therefore includes HTTP transport, queue and barrier wait, (de)serialization, and the graph and provenance bookkeeping. Failures and stragglers are injected at controlled rates from a queue model that approximates a remote QPU. Every configuration is run 10 times; we report medians with interquartile ranges, writing $p50$ for the median and $p99$ for the 99th-percentile tail.

\section{Results}
\label{sec:results}

\subsection{Orchestration Cost (RQ1)}
\label{sec:rq1}
Over 120 managed runs the median client-side end-to-end time was 783.18~s, of which 484.08~s (61.8\%) was orchestration time in the sense of Sec.~\ref{sec:eval}. That covers HTTP transport, queue and barrier wait, (de)serialization, and the graph and provenance bookkeeping. The per-round barrier waits in Table~\ref{tab:overheadpat} make clear that most of it is waiting, not computation inside the layer. The aggregate also hides a large difference between the two patterns.

Grouping runs by their recorded \texttt{pattern} facet (Table~\ref{tab:overheadpat}) separates them. Runs from before the facet existed are left out of the split, which is why the job counts in the table add up to fewer than 120. The iterative-consensus pattern pays for orchestration every round: a fan-out, a barrier, a consensus/dual update, repeated until the residual test passes. Its share, 78.4\%, is almost entirely the recurring barrier wait, and it grows with the number of iterations. The hierarchical pattern pays mostly once, for a single shallow fan-out and a stitch/feasibility check, with a bounded re-dispatch only when a cluster turns out infeasible; its 6.2\% is mostly the stitch. The two patterns ran on different services and instances, so this is a characterization of each, not a ranking. The practical reading is that the iterative pattern is the one a scheduler can actually speed up, by cutting the number of times a round has to cross to the device and back; the hierarchical pattern barely notices that boundary.

\begin{table}[t]
\caption{Orchestration cost by fan-out pattern (grouped on the recorded \texttt{pattern} facet). ``B/s'' is the median per-round barrier wait (iterative) or stitch/refine wait (hierarchical).}
\label{tab:overheadpat}
\centering
\small
\begin{tabular}{@{}p{2.55cm}p{0.55cm}p{0.95cm}p{0.95cm}p{1.7cm}@{}}
\toprule
\textbf{Pattern} & \textbf{Jobs} & \textbf{Ovh.} & \textbf{B/s (ms)} & \textbf{Dominant comp.} \\
\midrule
Iterative (ADMM) & 81 & 78.4\% & 36980 & per-round barrier \\
Hierarchical (MDCVRP) & 33 & 6.2\% & 92888 & stitch/refine \\
\bottomrule
\end{tabular}
\end{table}

These figures are for the managed engine with provenance switched on. Each subproblem adds one short append to a log on the shared filesystem, so provenance storage grows with the number of subproblems and iterations and does not depend on the backend. We have not yet measured the same runs with provenance off or on the unmanaged driver script; that comparison is under way and will go into a revised version.


\subsection{Feedback-Driven Termination (Provenance View)}
\begin{figure}[t]
\centering
\includegraphics[width=\linewidth]{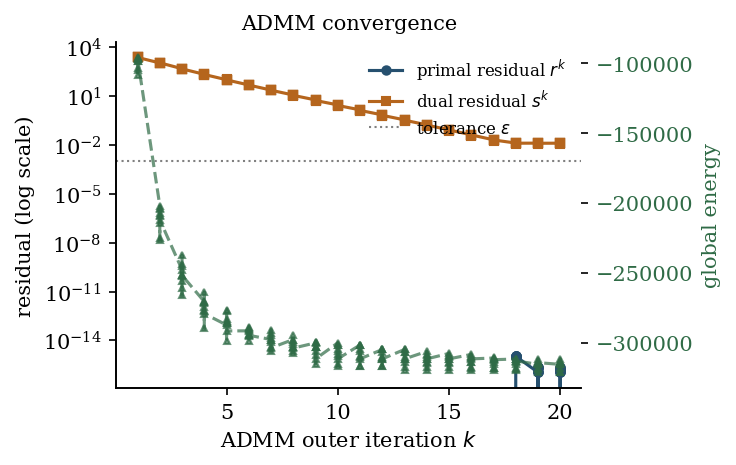}
\caption{Feedback-driven termination as captured provenance. The recorded primal/dual residual trajectory closes the dynamic graph: when both residuals fall below tolerance $\epsilon$ the workflow stops generating fan-outs and the run is classified converged.}
\label{fig:admm}
\end{figure}

Whether the workflow emits another fan-out or stops is decided by the residual trajectory in the \texttt{IterationRecord}s. Fig.~\ref{fig:admm} plots the primal and dual residuals of one ADMM run, rebuilt from those records rather than from anything inside the solver. The point where both drop below the tolerance $\epsilon$ is the point where the graph stops growing, and it can be checked after the fact.

\subsection{Resilience Under Injected Failures (RQ2)}
\label{sec:rq2}
\begin{figure}[t]
\centering
\includegraphics[width=\linewidth]{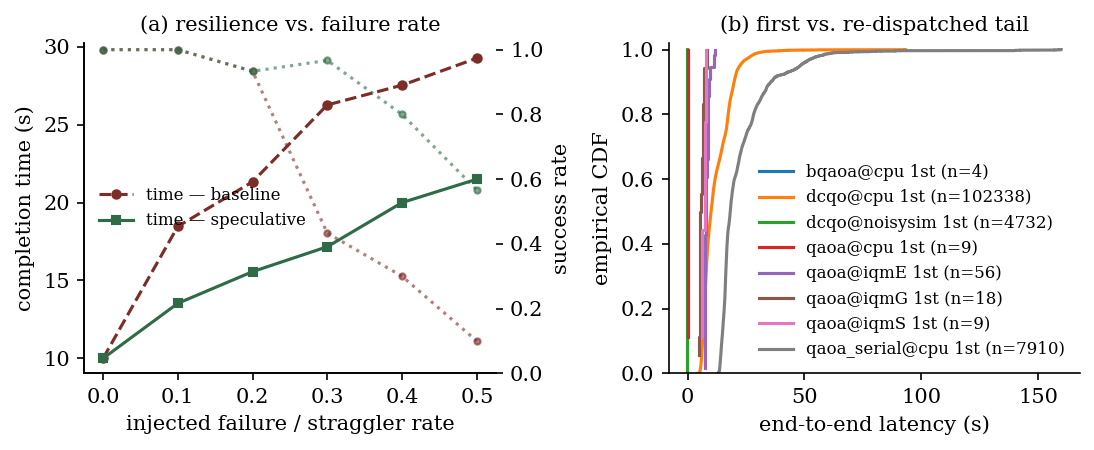}
\caption{Workflow resilience under injected failures. (a) As the injected failure/straggler rate increases, speculative re-execution sustains completion where the unmanaged driver stalls. (b) Separating first-attempt from re-dispatched subproblems by attempt number shows that bounded retries absorb stragglers in a distinct tail without shifting the first-attempt distribution that sets steady-state iteration makespan.}
\label{fig:resilience}
\end{figure}

Two different regimes need to be kept apart here. In \emph{production} the engine logged 4736 failed or timed-out blocks, yet not one was re-dispatched and no speculative copy was ever launched. At the failure rates we actually saw, quorum-deferred aggregation and warm-starting absorbed every loss before the reschedule path was reached, so the steady-state latency tail is set entirely by first attempts. That is why we ran the \emph{injected-failure sweep} of Fig.~\ref{fig:resilience}: production never exercised the recovery path, and we wanted to see it work.

Panel~(a) shows that as the injected rate goes up, speculative re-execution keeps runs completing where an unmanaged driver, the same loop without the layer, simply stalls. Panel~(b) uses the attempt number in each \texttt{BlockRecord} to split first attempts from re-dispatched work. Bounded re-dispatch picks up stragglers and failed submissions in a tail of its own, and the first-attempt distribution, which is what sets a typical iteration's makespan, does not move. We can only draw this picture because recovery actions are recorded as provenance: the attempt counter that drives idempotent re-execution (R-C) is the same field that tells us whether a slow record is slow because of the backend or because it was retried, and it is recorded the same way for both patterns.

\subsection{Backend Portability and Real-QPU Reach (RQ3)}
\label{sec:rq3}
Each \texttt{BlockRecord} carries the backend descriptor that was resolved at solve time: provider, device type, device name and shot budget. So the same provenance that gave us the orchestration numbers also tells us which sampler ran each subproblem, and how long it took and how many shots it used.

Table~\ref{tab:backend} breaks the runs down by device. Of the 8 backends used, 3 were QPUs, among them IQM Emerald and Garnet (\texttt{iqmE}, \texttt{iqmG}) at 8192 shots per circuit. The per-device tails put the simulator-versus-QPU gap in plain numbers, and they are what a placement or sharing policy needs in order to judge when a submission is worth its queue. Recording the backend per subproblem rather than per run is what makes this possible under failover, where different subproblems of one iteration may end up on different samplers.

\begin{table}[t]
\caption{Per-backend breakdown over subproblem records: device type, blocks run, end-to-end latency tail and shot budget, from the device descriptor stamped into each record.}
\label{tab:backend}
\centering
\small
\begin{tabular}{@{}p{2.2cm}p{1.0cm}p{0.6cm}p{0.95cm}p{0.95cm}p{0.7cm}@{}}
\toprule
\textbf{Device} & \textbf{Type} & \textbf{Blk} & \textbf{p50\,ms} & \textbf{p99\,ms} & \textbf{Shots} \\
\midrule
dcqo@cpu & Simulator & 102338 & 10081 & 30092 & 256 \\
qaoa@cpu & Simulator & 7919 & 18292 & 62080 & 2000 \\
qaoa@iqmE & QPU & 56 & 7787 & 11935 & 8192 \\
qaoa@iqmG & QPU & 18 & 5487 & 8170 & 8192 \\
\bottomrule
\end{tabular}
\end{table}

\section{Discussion and Limitations}
\label{sec:discussion}
Running the whole loop on HPC brings out problems that single-device solver experiments never show: the volume of provenance at high fan-out, contention at aggregation, and how much completion depends on the straggler policy. All three stay or get worse once a queued QPU is one of the backends. Fixing them on HPC first is, in our view, what quantum readiness actually consists of.

The pattern characterization has a direct consequence for resource management. The iterative pattern spends most of its time at the per-round barrier, and how many times a round crosses to the device is fixed by the block solver's execution mode: one batched submission per round for a non-variational solver, one submission per optimizer iteration per block for a variational one. That choice, rather than solver quality, is the knob a scheduler above QPU-sharing middleware such as \cite{cipollini2026share} should be turning. We look at it in separate work.

Nothing here says which solver or backend gives better solutions. That is a benchmarking question, and a different paper.

Some limitations of our work need to be mentioned. Although some subproblems ran on IQM QPUs, most of the orchestration evaluation is HPC rehearsal with an injection model, which does not reproduce real queue dynamics at scale. The orchestration cost is reported for the managed engine only; the next experiment is a failure-free comparison against the unmanaged driver, against a provenance-off configuration, and against the loop rewritten in a generic WMS such as PyCOMPSs or Parsl. Nested patterns are supported but not evaluated. A proper multi-scheduler portability study and a scalability study across fan-out are still to be done. And the readiness levels are a way of organizing the work, not a validated metric.

\section{Conclusion}
\label{sec:conclusion}
Iterative hybrid quantum--classical decomposition deserves to be run as a scientific workflow, and HPC is the place to build and test that workflow before fault-tolerant QPUs arrive. We looked at two decomposition patterns taken from real applications and found that they differ by an order of magnitude in what they ask of the orchestration layer, with the iterative one dominated by a barrier that recurs every round. We described a workflow layer that gives a generic engine what this loop needs and the engine does not have: a termination node driven by feedback, warm-start and quorum recovery, backend failover within a round, and a provenance schema stamped with device and shot budget. We measured what the layer costs on our engine, showed that it keeps runs completing under injected failures that stall an unmanaged driver, and pulled per-device latency tails for simulators, emulators and IQM QPUs out of the same provenance. Next are the baselines against the unmanaged driver and against a generic WMS, an evaluation of nested patterns, more real-QPU execution, and enactment across facilities.

\section*{Acknowledgments}
This work was supported in part by the Italian Ministry of Environment and Energy Security (MASE) through the GENESIS project (ID MI DDR 00308) under the Mission Innovation 2.0 initiative (DM n. 386/2023), the Sardinia Regional Authorities, and the Italian Ministry of Enterprises and Made in Italy (MIMIT) via the 5G technology support program under Axis 1 'House of Emerging Technologies' (CTE) for the 'Cagliari Digital Lab' project (ID: G27F22000040008).

The authors acknowledge IQM Quantum Computers for awarding computational credits that enabled access to real quantum hardware used in this work.

The authors acknowledge the use of Anthropic's Claude, a large language model, to assist with code structure, debugging, and language editing. All AI-assisted content was reviewed, edited and verified by the authors.

\bibliographystyle{IEEEtran}
\bibliography{workflow}
\end{document}